# Why Meditation Wearables Fail: Reward Misspecification in Closed-Loop EEG and Biofeedback Systems

**Joy Bose**
Independent Researcher, Bengaluru, India
joy.bose@ieee.org

**Abstract**

Consumer EEG headbands, HRV biofeedback devices, and closed-loop neurostimulation systems are a growing class of wearable technology marketed for meditation enhancement and mental wellbeing. These systems monitor physiological signals in real time and deliver feedback intended to guide users toward target brain or body states. We argue that all current systems in this class share a fundamental design flaw: they reward measurable proxy signals rather than the outcomes they claim to produce. When a user optimises for calm EEG, HRV coherence, or breathing resonance, their brain learns to produce those signals through whatever strategy is most efficient, including strategies that have nothing to do with the intended benefit. We formalise this as reward misspecification: the policy that maximises the proxy reward R_proxy is not the policy that maximises the true intended outcome V_target (arg max_pi E_pi[R_proxy] is not approximately equal to arg max_pi E_pi[V_target]). This leads to three predictable failure modes: proxy mismatch, strategy shortcutting, and transfer failure. We review how existing devices including the Muse EEG headband, HeartMath HRV systems, Unyte IOM2, and clinical neurofeedback systems instantiate these failures. We introduce a four-tier measurability taxonomy that distinguishes wearable targets that are reliably measurable with current technology (Tier 1) from targets that are currently unmeasurable (Tier 3) or possibly structurally unmeasurable in third-person terms (Tier 4), and we show that most current devices make implicit Tier 3 and Tier 4 claims. We then propose a design framework that avoids the three failure modes: single Tier-1 target (mind-wandering onset detected from EEG), negative-only cueing with no proxy reward, temporal separation of fast EEG and slow somatic feature streams, and transfer to unassisted practice as the only success criterion. No current product meets all four criteria. The framework has direct implications for the design, evaluation, and regulation of the next generation of cognitive and contemplative wearables.

## 1. Introduction

Wearable EEG and biofeedback devices for meditation and mental wellbeing have grown from academic curiosities into a commercial category. The Muse S headband (InteraXon) measures frontal and temporal EEG and plays audio feedback correlated with estimated calm states. The HeartMath Inner Balance monitors heart rate variability and scores breathing coherence. The Unyte IOM2 measures HRV via an ear clip and uses the signal to advance users through gamified breathing exercises. Research-grade systems add transcranial direct current stimulation (tDCS), transcranial alternating current stimulation (tACS), and functional near-infrared spectroscopy (fNIRS) to the sensor stack. Clinical neurofeedback systems used in ADHD and trauma treatment reward specific EEG frequency bands using continuous audio or visual feedback.

All of these systems share the same basic architecture. A sensor measures a physiological signal. An algorithm classifies the signal as good or bad relative to a target. The system delivers feedback that rewards the good state. The user, receiving this feedback over many sessions, learns to produce the good state more reliably.

The problem is that the good state as defined by the sensor is not the same as the benefit the device claims to produce. Calm EEG can be produced by genuine settled awareness, mild dissociation, pleasant drowsiness, or a jaw position that creates the target waveform as an artefact. HRV coherence goes up when you are genuinely calm, but also when you are in early-stage hypnagogic sleep or mildly dissociated. A breathing resonance score goes up when you breathe at the right rate, but the score says nothing about whether you are mentally present or mentally elsewhere.

In machine learning, this is called Goodhart's Law: when a measure becomes a target, it ceases to be a good measure (Goodhart, 1984). In AI alignment research, the same problem is called reward misspecification: you specify a proxy for what you want, the system optimises the proxy, and you end up with something that scores well on the metric while failing to do what you actually wanted (Turner et al., 2020, Russell, 2022).

We argue that every current meditation wearable and closed-loop neurofeedback system suffers from reward misspecification. This is not a criticism of any specific product. It is a structural problem that follows from the combination of an adaptive user and a proxy reward signal, and it will affect any system that rewards a physiological measurement as a proxy for a psychological outcome.

The history of contemplative neurotechnology includes earlier cautionary examples. The God Helmet, developed by Persinger and colleagues in the 1980s, applied weak magnetic fields to the temporal lobes and produced reports of anomalous experiences in many subjects (Persinger & Valliant, 1985). Subsequent work found that these effects were better predicted by participant suggestibility than by the magnetic field application itself (Granqvist et al., 2005). The lesson is not simply that the device failed to replicate, but that it was targeting an outcome, i.e. anomalous experience, that the contemplative traditions it claimed to engage treat as a peripheral and potentially misleading artefact of practice. Misspecifying the target is not a new problem in this field.

This paper makes three contributions. First, we identify and name the three specific failure modes that reward misspecification produces in wearable EEG and biofeedback systems: proxy mismatch, strategy shortcutting, and transfer failure. Second, we review how existing devices instantiate each failure mode. Third, we propose a design framework that avoids all three, grounded in existing EEG and neurofeedback literature, and we show what a device built to this framework would look like.

## 2. How Current Devices Work and Where They Fail

### 2.1 The Standard Architecture

Every closed-loop wearable operates through four stages. Sensing: a physiological signal is measured continuously. Classification: an algorithm determines whether the current state is on-target or off-target. Feedback: the device delivers a signal (audio, visual, haptic) contingent on the classification. Learning: the user, over many sessions, adjusts their behaviour in response to the feedback.

This architecture is not inherently flawed. Neurofeedback training for ADHD uses exactly this architecture, and there is reasonable evidence that it produces genuine changes in attentional control (Ros et al., 2020). The problem emerges specifically when the feedback signal is a proxy for a psychological outcome that is unmeasurable directly, and when the outcome is complex enough that many different strategies can produce the proxy signal.

Meditation is the worst-case scenario for this architecture. The intended outcome, something like reduced reactivity, improved metacognitive awareness, or reduction of habitual distress, cannot be directly measured by any current wearable. The proxy signals available (EEG band power, HRV, skin conductance, respiration rate) all have multiple causes, most of which have nothing to do with the intended outcome. And the human brain is an excellent optimiser, capable of discovering strategies to produce any reliable feedback signal that a session of practice provides time to discover.

### 2.2 Failure Mode 1: Proxy Mismatch

The first failure mode occurs when the proxy signal simply does not correspond to the intended outcome even in principle.

Consider the Muse headband. Muse classifies EEG states as calm or active using primarily frontal and temporal channels, and plays soothing audio when calm is detected and rain or storm sounds when activity is detected. The intended outcome is presumably improved meditation quality. But EEG calm as Muse defines it is elevated in drowsy, dissociated, and passively relaxed states as well as in genuinely meditative states. There is no evidence that users who score well on the Muse calm metric have better meditation outcomes than users who do not. The proxy does not correspond to the target.

HeartMath HRV systems have the same problem at the physiological level. HRV coherence, measured as the ratio of heart rate variability power at the breathing resonance frequency, is a marker of parasympathetic tone. Parasympathetic activation correlates with reduced stress in many contexts. But it also occurs during early sleep, passive relaxation, and certain dissociative states. The claim that rewarding HRV coherence produces equanimity is not supported by the available evidence.

The Unyte IOM2, which the author has used personally, closes the biofeedback loop more completely than most systems: the game advances when you breathe slowly, and you receive direct feedback that your breathing is producing the intended physiological response. This is a well-designed HRV biofeedback system. However, it still rewards a proxy (breathing resonance score) that is at best a correlate of relaxation, not a measure of the contemplative skill the system is marketed to develop. A user can achieve high resonance scores while mentally planning their shopping list. The IOM2 has no way to know the difference.

Clinical neurofeedback systems reward specific EEG frequency ratios, most commonly increasing theta relative to beta for ADHD or increasing alpha for anxiety and attention training. These systems have more evidence behind them than consumer devices, and the evidence is better controlled. However, even here, the mechanism is poorly understood. Subjects trained to increase frontal midline theta do so, but the strategy they use to produce the target waveform varies across individuals and does not always correspond to what researchers intended (Ros et al., 2020). Some increase the target frequency through genuine attentional regulation. Others use postural or facial strategies that produce the frequency as a signal artefact.

### 2.3 Failure Mode 2: Strategy Shortcutting

The second failure mode occurs when the user learns a strategy that produces the proxy signal without the intended underlying process.

This is the most insidious failure because it is invisible both to the device and to the user. The device sees the target signal and provides positive feedback. The user experiences positive feedback and feels they are making progress. Neither has any way to know that the strategy producing the signal is not the strategy that produces the intended benefit.

**Examples of strategy shortcutting in EEG-based systems include:** Orofacial strategies. Small jaw movements, changes in eye position, and facial tension produce broadband EEG artefacts that can increase or decrease apparent band power in target frequency ranges. Users who discover that a particular facial configuration reliably produces the target state have found a shortcut that exploits the sensor rather than developing the intended skill.

**Postural strategies.** Certain head positions and body postures affect EEG signal quality and band power measurements. A user who learns that sitting very still with their head in a particular orientation consistently produces a calm classification has learned a postural trick, not a meditative one.

**Breathing strategies.** Paced breathing at specific rates affects both EEG and HRV in ways that can produce high scores on biofeedback systems. A user who learns to breathe at 6 breaths per minute consistently achieves good HRV coherence. Whether they have learned genuine breath-body coordination or merely a breathing rate is not determinable from the sensor data.

**Suppression**. Perhaps the most concerning shortcut is emotional suppression. Suppression actively dampens physiological arousal markers. A user who learns to suppress emotional responses will show reduced galvanic skin response, lower HRV low-frequency power, and quieter EEG relative to an active emotional response. Biofeedback systems that reward these signals will reward suppression as reliably as they reward genuine equanimity. Research on adverse effects of intensive meditation practice has documented cases where practitioners who believed they were developing equanimity were in fact engaging suppression, with significant negative consequences over time (Britton et al., 2014).

The key point is that strategy shortcutting is not user dishonesty. It is the natural consequence of placing an optimising system (the user's brain) inside a feedback loop defined by a proxy metric. The brain will find the most efficient path to the reward. If the most efficient path involves a genuine skill, the device works. If the most efficient path involves a shortcut, the device fails while appearing to succeed.

### 2.4 Failure Mode 3: Transfer Failure

The third failure mode occurs when skills developed with device assistance do not transfer to unassisted contexts.

This is the most straightforward failure to understand and the most completely ignored by the current market. No consumer meditation wearable currently on the market requires users to demonstrate that their in-session gains persist in unassisted practice. No evaluation framework for clinical neurofeedback makes transfer to unassisted function the primary outcome. Sessions are evaluated by in-session metrics, and success is defined as improvement on those metrics.

This matters because the entire rationale for a meditation wearable is that it accelerates skill development. A device that produces good in-session scores but does not transfer has not accelerated skill development. It has created a dependency. The user is calm when the device is on and unchanged when the device is off. This is the opposite of what was intended.

Transfer failure is related to but distinct from the other failure modes. A device might accurately target the right proxy, users might learn the right strategy, and the skill still might not transfer if it was acquired in conditions too narrow to generalise. Meditation practice has always recognised this problem. The insistence in many traditions on practising in a variety of conditions, on applying insight in ordinary life as well as on the cushion, and on gradually withdrawing external support as the practitioner develops, all reflect an understanding that skills trained only in idealised conditions do not automatically apply in other conditions. Device-assisted practice creates exactly these idealised conditions and then removes the device, expecting the skill to have generalised.

## 3. What the Evidence Actually Shows

### 3.1 EEG Neurofeedback

The neurofeedback literature is more developed than the consumer wearable literature and shows more rigorous results, but the same problems appear in a more careful form.

A meta-analysis by Arns and colleagues found that EEG neurofeedback for ADHD produced significant improvements in inattention and hyperactivity, but that the specificity of EEG changes as the mediating mechanism was not demonstrated (Arns et al., 2014). That is, users improved, but whether they improved because of the EEG changes targeted by the protocol or because of non-specific factors (engagement, attention to internal states, expectation of benefit) was unclear.

Frontal midline theta neurofeedback has been used to train focused attention states relevant to meditation. Brandmeyer and Delorme demonstrated that closed-loop frontal midline theta training during focused attention practice produced EEG changes consistent with the training protocol (Brandmeyer & Delorme, 2020). This is a more careful study than most consumer device research, and its results are promising. However, transfer to unassisted practice was not the primary outcome, and the duration of training was short relative to what a meditator would typically undertake.

A systematic review of mindfulness-based neurofeedback found that within-session effects on targeted biomarkers were reasonably consistent but that transfer effects and mental health benefits remained inconsistently demonstrated, and that many studies lacked adequate sham controls (Treves et al., 2024). The pattern is consistent with the failure modes described here: devices produce in-session signal changes, but those changes do not reliably produce lasting benefits.

### 3.2 tDCS and Closed-Loop Stimulation

Transcranial direct current stimulation has been investigated as an adjunct to meditation and attention training. The evidence is mixed. Stimulation of the prefrontal cortex has been shown to modestly improve sustained attention in fatigued subjects, but effect sizes are small and highly variable across individuals (Bikson et al., 2019). The montage sensitivity of tDCS effects means that protocols that work in one lab context may not replicate in another.

More relevant to this paper is the interaction between tDCS and closed-loop EEG monitoring. Firing 1-2 mA direct current pulses introduces electromagnetic interference that can saturate high-gain biopotential amplifiers measuring microvolt-level EEG signals. Any system that combines tDCS with simultaneous EEG must implement temporal multiplexing: stimulate during one window, record during the next. Without this, the EEG signal during stimulation is corrupted by stimulation artefacts and cannot be used for classification. This is a known hardware problem that is addressed inconsistently in current research systems and essentially never addressed in consumer products.

Transcutaneous auricular vagus nerve stimulation (taVNS) is a more conservative stimulation approach that targets the autonomic nervous system rather than cortical firing thresholds directly. Stimulation of the vagal branch at the ear tragus activates the nucleus tractus solitarius in the brainstem, producing bottom-up parasympathetic upregulation. The alignment risk of taVNS is lower than tDCS because it affects the physiological substrate for practice rather than attempting to directly induce a cortical state. The evidence for autonomic effects is stronger than for cognitive effects of tDCS (Colzato et al., 2018). This makes taVNS a more defensible adjunct to meditation wearables than direct cortical stimulation, provided it is implemented with appropriate isolation from the EEG sensing electronics.

### 3.3 HRV and Respiratory Biofeedback

HRV biofeedback is the most evidence-based form of biofeedback for stress and wellbeing outcomes. Meta-analyses show consistent effects on HRV measures and moderate effects on stress and anxiety (Lehrer & Gevirtz, 2014). The cardiovascular resonant frequency protocol, which trains users to breathe at approximately 0.1 Hz (6 breaths per minute) to maximise HRV coherence, has the strongest evidence base in the HRV biofeedback literature.

However, the transfer question is underexplored even in this best-case literature. Most studies measure HRV outcomes in session or immediately after. The question of whether users who complete a course of HRV biofeedback maintain improved autonomic regulation in daily life without the device is addressed in some studies but not consistently. The studies that do address it show mixed results.

The consumer deployment of HRV biofeedback in devices like HeartMath and Unyte adds the proxy mismatch problem to the transfer problem. These devices reward breathing resonance as a proxy for wellbeing, and they do so with gamification (scores, advancement, points) that creates exactly the conditions for the strategy shortcutting failure mode described above.

### 3.4 A Taxonomy of What Wearables Can and Cannot Measure

A recurring theme in the analysis above is that different targets carry very different levels of measurability risk. Existing products conflate targets that are measurable with reasonable confidence with targets that are currently unmeasurable or possibly structurally unmeasurable in third-person terms. We propose a four-tier taxonomy to make these distinctions explicit (Table 1).

***Table 1.*** *Measurability taxonomy for closed-loop wearable targets. Tier 1 states are the only defensible primary targets for current devices. Tier 3 and Tier 4 claims by any device constitute a misrepresentation of the state of the evidence.*

| Tier | Description | Examples (device-familiar terms) | Design implication |
|---|---|---|---|
| 1 | Reliably measurable with current technology | Mind-wandering onset; gross agitation (elevated HRV LF/HF, GSR spike, movement burst); gross dullness (postural collapse, theta elevation, respiration slowing); sustained attention duration | Safe primary targets for closed-loop systems. Muse calm detection and HRV coherence scoring claim Tier 1 but lack validation for specificity. |
| 2 | Candidate correlates requiring validation | DMN suppression index; frontal midline theta power (FM-theta); gamma synchrony (Lutz et al., 2004) during open monitoring | Requires individual calibration and sham-controlled validation before clinical or commercial use. Confounds with artefact and non-specific arousal changes must be addressed. |
| 3 | Phenomenologically distinct but currently unmeasurable in third-person terms | Subtle laxity vs. genuine stillness; equanimity vs. suppression; metacognitive clarity vs. passive drifting | Physiological proxies alone are insufficient. First-person report is essential. Behavioural assays (reaction time, emotional response flexibility) may eventually provide partial Tier 2 access. Contemplative traditions provide the most detailed existing accounts of these distinctions: the Theravada vipassana literature on the corruptions of insight (Mahasi Sayadaw, 1978) and Tibetan Mahamudra and Dzogchen accounts of laxity versus genuine stillness (Longchenpa, 2001) predate third-person |

| Tier | Description | Examples (device-familiar terms) | Design implication |
|---|---|---|---|
| | | | neuroscience by centuries and constitute an independent empirical resource for calibrating Tier 3 distinctions. No current wearable can distinguish these states. |
| 4 | Possibly structurally unmeasurable in third-person terms | Enlightenment; nondual awareness; rigpa recognition; stream-entry; liberation | No device claim is possible. Marketing language claiming to accelerate, induce, or measure these states is a misrepresentation regardless of the sensor quality. |

This taxonomy has direct regulatory and consumer protection implications. Devices currently marketed for equanimity training (Tier 3), stress transformation (Tier 3 to 4), or consciousness expansion (Tier 4) are making claims that go well beyond what any wearable sensor can validate. The design framework in Section 4 is explicitly restricted to Tier 1 targets.

## 4. A Design Framework That Avoids These Failures

### 4.1 Four Design Criteria

We formalise the core problem before specifying the design response.

Let $R_{\text{proxy}}$ denote the device's measured reward signal (for example, calm EEG power, HRV coherence, or breathing resonance score), and let $V_{\text{target}}$ denote the true intended outcome (for example, reduced distractibility, improved emotional regulation, or reduced habitual distress). The design failure of all current meditation wearables is that the policy a user learns to maximise R_proxy is not the policy that maximises V_target. The problem (which we call the contemplative alignment problem) arises when:

$$\arg\max_{\pi} \mathbb{E}_{\pi}\left[R_{\text{proxy}}\right] \not\approx \arg\max_{\pi} \mathbb{E}_{\pi}\left[V_{\text{target}}\right]$$

That is, the policy $\pi$ that maximises the proxy does not correspond to the policy that maximises the true value. The practitioner's brain, as an adaptive optimiser, will converge on arg max$\pi$ $R_{\text{proxy}}$. If the gap between $R_{\text{proxy}}$ and $V_{\text{target}}$ is non-trivial, the result is a practitioner who scores well on the device's metric while failing to develop the intended capacity, and who may be unaware that this is occurring.

The user's brain, as an adaptive optimiser, converges on the left side. All three failure modes described in Section 2 follow directly from this gap. Proxy mismatch arises because R_proxy is the wrong function. Strategy shortcutting arises because the user finds efficient paths to R_proxy that bypass the mechanism intended to produce V_target. Transfer failure arises because the policy learned to maximise R_proxy in the training context does not generalise to contexts where R_proxy is unavailable.

Based on the analysis above, we propose four design criteria for closed-loop meditation wearables that avoid the three failure modes.

- **Criterion 1: Single measurable target.** The system must target one physiological state that has a validated neurophysiological correlate and that precedes the intended outcome in the causal chain. For meditation, the most defensible single target is the onset of mind-wandering. EEG markers of default mode network activation reliably precede conscious recognition of distraction by several seconds (Smallwood & Schooler, 2006). This is the strongest Tier-1 measurable state in contemplative neuroscience: it has cross-study replication, clear mechanistic basis, and an obvious intervention point. All other targets (equanimity, compassion, insight, nondual awareness) are either unmeasurable with current wearable technology or have insufficient evidence for the claimed signal-outcome relationship.

- **Criterion 2: Negative-only feedback.** The system must deliver a cue only when the target state (distraction onset) is detected. It must never deliver a reward signal for achieving a positive state. No pleasant sounds for calm EEG. No scores or resonance ratings. No progress bars or achievement badges. The cue is neutral: a brief bone-conduction tone or ambient acoustic shift. The only information it conveys is: distraction has been detected. This avoids the proxy mismatch failure because nothing is being rewarded as a proxy for the intended outcome. It avoids the strategy shortcutting failure because there is no reward signal to shortcut toward.
- **Criterion 3: Layer separation by temporal scale.** EEG-based mind-wandering detection must use only high-speed features computed on 500 millisecond windows. Heart rate variability, respiration, and inertial measurement unit data have physiological and computational latency of several seconds and cannot capture the micro-moment of distraction onset. If slow metrics are included in the Layer 1 classifier, cues will fire seconds after the user has already drifted into a narrative loop. HRV, respiration, and posture data can be used for gross state estimation (detecting agitation or dullness) but must not participate in the fast distraction detection loop.
- **Criterion 4: Transfer to unassisted practice as the primary outcome.** The system must measure the correction interval (distraction onset to conscious self-correction) in device-absent sessions throughout and after training. If this interval does not decrease in unassisted sessions, the system has produced dependency rather than skill and has failed regardless of in-session performance. This criterion makes the system's success claim falsifiable and prevents the transfer failure mode.

### 4.2 Hardware Stack

A system meeting these criteria requires the following minimum hardware:

- **EEG**: 4 to 8 dry electrodes at frontal and temporoparietal sites (Fp1, Fp2, Fz, Cz, TP9, TP10). Consumer options include the Muse S headband with BrainFlow API access, which provides four channels adequate for frontal alpha asymmetry and frontal midline theta computation. Research options include the OpenBCI Cyton, which provides 8 channels at 24-bit resolution using the Texas Instruments ADS1299 analog front-end.
- **HRV**: A photoplethysmography sensor at the ear or finger for inter-beat interval extraction. The MAX30102 breakout board interfaces via I2C to a microcontroller and costs approximately USD 9. HRV is used for gross autonomic state monitoring (Layers 2 and 3) but not for Layer 1 distraction detection.
- **Respiration**: A conductive stretch sensor on a chest strap for breath rate and pattern monitoring. This provides a slow-timescale context signal for autonomic support but does not participate in Layer 1 classification.
- **IMU**: A 9-axis inertial measurement unit (BNO085 or MPU6050) for posture monitoring. Postural collapse combined with elevated theta and slowed respiration is a reasonable marker of gross dullness. This is a Layer 2 signal.
- **Bone conduction transducer**: For cue delivery without blocking the ear canal. The Dayton Audio BCT-2 costs approximately USD 13 and is driven by a MAX98357A I2S amplifier.
- **Microcontroller**: ESP32-S3 for sensor aggregation, BLE streaming to a phone or laptop, and cue delivery.

Total component cost for the entry-level prototype (Muse S plus ESP32 auxiliary board) is approximately USD 450. The research-grade OpenBCI path costs approximately USD 1,300.

### 4.3 Signal Processing Architecture

The signal pipeline separates into two temporally distinct streams that must not be merged for Layer 1 classification.

The fast neural path uses EEG-only features on 500 millisecond windows stepped at 250 milliseconds. Features include frontal alpha asymmetry, fronto-central phase-locking value in the theta band (4 to 8 Hz; Aftanas & Golocheikine, 2001), theta-to-beta ratio at Fz and Cz, and Hjorth mobility and complexity measures. These features are computed per window and fed to a personalised binary classifier (Random Forest or XGBoost) trained on the individual user's Phase A calibration data.

The slow somatic context path uses HRV (RMSSD, LF/HF ratio), respiration (rate, depth, irregularity), and IMU (head pitch, movement jitter) on 10 to 30 second sliding windows. These features inform Layer 2 state estimation (gross agitation and dullness) and Layer 3 taVNS triggering, but they do not participate in Layer 1 cue logic.

The personalised classifier is trained during a 2-week passive sensing baseline (Phase A). At random intervals during baseline meditation sessions, the system prompts the user for a phenomenological self-report: settled, wandering, or unclear. This approach draws on neurophenomenological methodology, which proposes that first-person reports constitute an independent source of evidence that can be coordinated with third-person physiological measurement rather than treated as noise (Varela & Shear, 1999; Varela et al., 1991). These labels, timestamped against the EEG feature vector, form the training set. A classifier trained on individual-specific data substantially reduces the proxy mismatch problem by calibrating the measurement to the specific user rather than relying on population-level priors.

### 4.4 The Transfer Test

The transfer test is the empirical procedure that makes the system's success claim falsifiable. It runs as follows.

- During Phase B (weeks 3 to 8), one session per week is conducted without any cues delivered. EEG is recorded passively. The correction interval (distraction onset detected by EEG to self-correction, estimated from subsequent probe responses) is logged.
- During Phase C (weeks 9 to 10), the device is physically removed. The user meditates without any sensing or feedback. The correction interval is estimated using app-timed phenomenological prompts only.

The system succeeds if and only if the Phase C correction interval is significantly shorter than the Phase A baseline. If the interval is shorter only in Phase B device-present sessions, the device has produced dependency and the experiment has failed. This is a straightforward pre-registered hypothesis that any research study using this system should state explicitly before data collection.

### 4.5 Comparison with Existing Products

Table 2 compares existing systems against the four design criteria.

| System | Single measurable target | Negative-only feedback | Layer separation | Transfer test |
|---|---|---|---|---|
| Muse S | No (rewards calm EEG broadly) | No (rewards calm) | No | No |
| HeartMath | No (rewards HRV coherence) | No (rewards coherence score) | Not applicable | No |

| System | Single measurable target | Negative-only feedback | Layer separation | Transfer test |
|---|---|---|---|---|
| Unyte IOM2 | No (rewards breathing resonance) | No (gamified reward) | Not applicable | No |
| Clinical neurofeedback | Partial (targets specific band) | No (rewards target band) | Rarely | Rarely |
| OpenBCI (raw) | N/A (no intervention) | N/A | N/A | N/A |
| Proposed system | Yes (mind-wandering onset only) | Yes | Yes | Yes (primary outcome) |

No current consumer or clinical system meets all four criteria.

Figure 1. The reward misspecification loop in closed-loop meditation wearables.

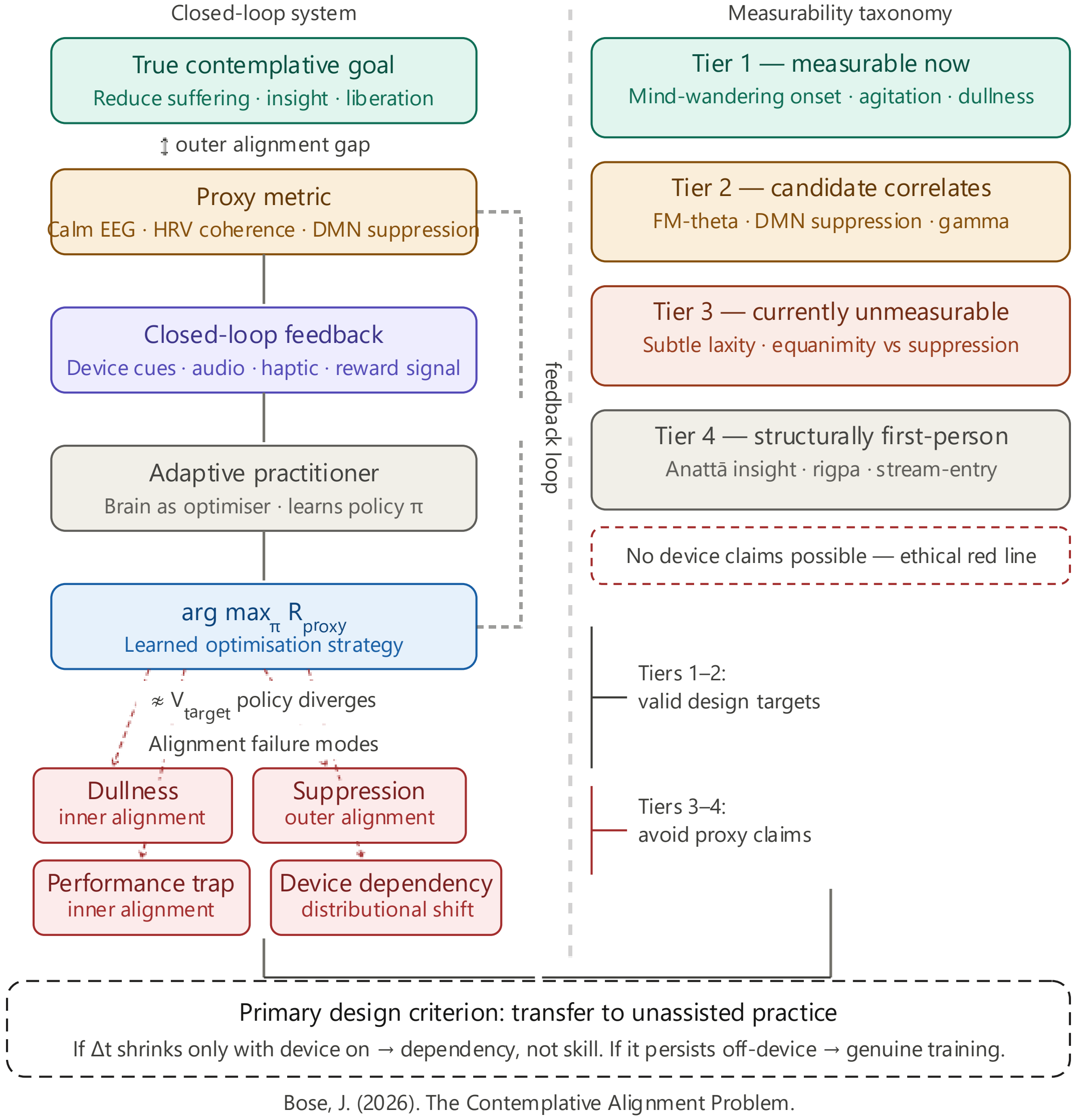

***Figure 1.*** *The Contemplative Alignment Problem: The closed-loop meditation wearable creates an optimisation loop in which the user's brain converges on arg max R_proxy (the device's proxy metric) rather than V_target (the true intended outcome). The right column shows the four-tier measurability taxonomy: only Tier 1 states are suitable primary targets for current devices. The transfer test at the bottom is the only outcome measure that distinguishes genuine skill development from device dependency.*

## 5. Hardware Safety for Combined EEG and Stimulation Systems

Any system that combines EEG sensing with electrical stimulation (tDCS, tACS, or taVNS) must address three hardware safety requirements that are consistently underspecified in published literature and absent from consumer products.

- **Electrical isolation.** The stimulation current generation circuit must run on a physically separate battery rail from the EEG sensing electronics. All digital control lines between the stimulator and the microcontroller must pass through optocouplers. This prevents stimulation switching noise from contaminating the EEG signal and eliminates the possibility of leakage current through sensing electrodes contacting the scalp.
- **Manual current setting.** Stimulation amplitude must be set by a physical potentiometer or hardware-level digital-to-analog converter at session start. Software must not adjust stimulation amplitude in real time in response to EEG classifier output. This prevents a closed-loop runaway condition in which the stimulator is optimised against the classifier's proxy signal, which is the stimulation equivalent of the reward misspecification problem.
- **Time-division multiplexing for EEG/stimulation co-acquisition.** When taVNS or tDCS pulses are active, electromagnetic interference saturates the high-gain analog front-end used for EEG. The firmware must implement temporal multiplexing: stimulate for 400 milliseconds, then pause for 100 milliseconds, recording EEG only in the silent window after a settling period of at least 50 milliseconds. EEG feature extraction must be restricted to windows that do not overlap with active stimulation periods.

## 6. Discussion

### 6.1 Implications for Product Design

The analysis in this paper implies that the core problem with current meditation wearables is not sensor quality, algorithm sophistication, or form factor. It is reward specification. A system that specifies the wrong reward will produce the wrong behaviour no matter how good its sensors are. The Muse S has adequate EEG quality for frontal alpha asymmetry computation. HeartMath has adequate HRV quality for cardiovascular resonant frequency training. The problem in both cases is not the signal quality. It is that the signal is being rewarded as an end in itself rather than being used to detect a specific cognitive event that is causally upstream of the intended outcome.

The proposed framework does not require better sensors. It requires a different reward structure. The only change that cannot be achieved with existing consumer hardware is the transfer test, which requires a study design commitment rather than a hardware change.

### 6.2 Implications for Research Design

Research evaluating closed-loop meditation wearables should pre-register transfer to unassisted practice as the primary outcome before data collection. In-session signal changes should be reported as mechanistic checks, not as primary outcomes. Sham-controlled designs are required to separate the effects of the feedback from the effects of the wearable context (wearing a device, attending to internal states, expecting benefit).

The phenomenological self-report calibration approach proposed here, using user-generated labels during passive sensing to train personalised classifiers, opens a research programme in neurophenomenology applied to wearable systems. The quality of these labels is itself a research question: beginners systematically mislabel subtle states, and the classifier inherits these biases. Label quality, classifier performance, and transfer outcomes should all be reported in studies using this approach.

### 6.3 Limitations

The design framework proposed here addresses proxy mismatch, strategy shortcutting, and transfer failure at the system level. It does not address individual variability in EEG signatures, which is high enough that a personalised classifier trained on two weeks of data may still perform poorly for some users. It does not address the phenomenological calibration problem: users who cannot accurately report their mental states provide noisy labels that degrade classifier performance. And it does not address the possibility that even accurate detection and cueing of mind-wandering onset may not produce meaningful improvement in meditation outcomes if the underlying mechanisms of skill transfer are more complex than simple metacognitive practice.

These are empirical questions that can only be answered by studies designed with the transfer test as the primary outcome. The framework proposed here is necessary but not sufficient for a useful meditation wearable. Whether it is sufficient is an open question.

### 6.4 Regulatory Implications

Consumer EEG and biofeedback devices are currently sold as wellness products rather than medical devices in most jurisdictions, which substantially reduces the evidence standard required for market entry. The three failure modes described here would not be detectable by any current regulatory framework for wellness devices. A device that produces good in-session scores, avoids obvious adverse effects, and makes wellness claims rather than medical claims can be sold without demonstrating transfer to unassisted function.

This is a gap worth addressing. At minimum, evaluation frameworks for cognitive and contemplative wearables should require pre-registered transfer testing as a condition of any efficacy claim. Devices that claim to improve meditation, attention, or stress resilience should be required to demonstrate that these improvements persist outside the device context.

## 7. Conclusion

Current EEG and biofeedback wearables for meditation and mental wellbeing share a structural design flaw: they reward measurable proxy signals rather than the outcomes they claim to produce. This produces three predictable failure modes: targeting the wrong proxy (proxy mismatch), learning the wrong strategy to produce the proxy (strategy shortcutting), and failing to transfer skill outside the device context (transfer failure). These failures are not product-specific. They follow from the architecture that all current systems share.

The proposed framework avoids these failures through four design criteria: single measurable target (mind-wandering onset via EEG), negative-only feedback (no reward for proxy states), temporal separation of fast EEG and slow somatic features, and transfer to unassisted practice as the primary success criterion. No current consumer or clinical system meets all four criteria. Implementation of this framework in hardware is feasible with existing components at approximately USD 450 for a prototype.

The central design principle generalises beyond meditation wearables. Any closed-loop system that places an adaptive user inside a feedback loop defined by a physiological proxy will face these failure modes. Getting the reward structure right is more important than getting the sensors right.

## Competing Interests

The author declares no competing interests.

**Data Availability Statement**

No new data were generated or analysed in this study.

**References**

Aftanas, L. I., & Golocheikine, S. A. (2001). Human anterior and frontal midline theta and lower alpha reflect emotionally positive state and internalized attention: high-resolution EEG investigation of meditation. Neuroscience letters, 310(1), 57-60.

Bikson, M., Paulus, W., Esmaeilpour, Z., Kronberg, G., & Nitsche, M. A. (2019). Mechanisms of acute and after effects of transcranial direct current stimulation. In Practical guide to transcranial direct current stimulation: principles, procedures and applications (pp. 81-113). Cham: Springer International Publishing.

Brandmeyer, T., & Delorme, A. (2020). Closed-loop frontal midline θ neurofeedback: A novel approach for training focused-attention meditation. Frontiers in human neuroscience, 14, 246.

Brandmeyer, T., Delorme, A., & Wahbeh, H. (2019). The neuroscience of meditation: Classification, phenomenology, correlates, and mechanisms. Progress in Brain Research, 244, 1-29.

Britton, W. B., Lindahl, J. R., Cahn, B. R., Davis, J. H., & Goldman, R. E. (2014). Awakening is not a metaphor: The effects of Buddhist meditation practices on basic wakefulness. Annals of the New York Academy of Sciences, 1307, 64-81.

Colzato, L. S., Sellaro, R., van den Wildenberg, W. P., & Hommel, B. (2015). tDCS of medial prefrontal cortex does not enhance interpersonal trust. Journal of Psychophysiology.

Evans, O., Stuhlmüller, A., Salvatier, J., & Filan, D. (2018). Modeling agents with probabilistic programs. arXiv preprint arXiv:1809.10156.

Goodhart, C. A. (1984). Problems of monetary management: the UK experience. In Monetary theory and practice: The UK experience (pp. 91-121). London: Macmillan Education UK.

Granqvist, P., Fredrikson, M., Unge, P., Hagenfeldt, A., Valind, S., Larhammar, D., & Larsson, M. (2005). Sensed presence and mystical experiences are predicted by suggestibility, not by the application of transcranial weak complex magnetic fields. Neuroscience Letters, 379(1), 1-6.

Turner, A., Ratzlaff, N., & Tadepalli, P. (2020). Avoiding side effects in complex environments. Advances in Neural Information Processing Systems, 33, 21406-21415.

Lehrer, P. M., & Gevirtz, R. (2014). Heart rate variability biofeedback: How and why does it work? Frontiers in Psychology, 5, 756.

Longchenpa. (2001). The Precious Treasury of the Basic Space of Phenomena. Trans. Richard Barron. Padma Publishing.

Lutz, A., Greischar, L. L., Rawlings, N. B., Ricard, M., & Davidson, R. J. (2004). Long-term meditators self-induce high-amplitude gamma synchrony during mental practice. Proceedings of the National Academy of Sciences, 101(46), 16369-16373.

Mahasi Sayadaw. (1978). The Progress of Insight. Trans. Nyanaponika Thera. Buddhist Publication Society.

Persinger, M. A., & Valliant, P. M. (1985). Temporal lobe signs and reports of subjective paranormal experiences in normal populations: A replication. Perceptual and Motor Skills, 60(3), 903-909.

Ros, T., Enriquez-Geppert, S., Zotev, V., Young, K. D., Wood, G., Whitfield-Gabrieli, S., Wan, F., et al. (2020). Consensus on the reporting and experimental design of clinical and cognitive-behavioural neurofeedback studies (CRED-nf checklist). Brain, 143(6), 1674-1685.

Russell, S. (2022). Human-Compatible Artificial Intelligence. Human-like machine intelligence, 1, 3-22.

Smallwood, J., & Schooler, J. W. (2006). The restless mind. Psychological Bulletin, 132(6), 946-958.

Sugiyama, M., & Kawanabe, M. (2012). Machine learning in non-stationary environments: Introduction to covariate shift adaptation. MIT press.

Treves, I. N., Greene, K. D., Bajwa, Z., Wool, E., Kim, N., Bauer, C. C., ... & Auerbach, R. P. (2024). Mindfulness-based neurofeedback: A systematic review of EEG and fMRI studies. Imaging Neuroscience, 2, imag-2.

Varela, F. J., Thompson, E., & Rosch, E. (1991). The Embodied Mind: Cognitive Science and Human Experience. MIT Press.

Varela, F. J., & Shear, J. (Eds.) (1999). The View from Within: First-Person Approaches to the Study of Consciousness. Imprint Academic.

Zander, T. O., & Kothe, C. (2011). Towards passive brain–computer interfaces: applying brain–computer interface technology to human–machine systems in general. Journal of neural engineering, 8(2), 025005.